\documentclass[preprint,12pt]{elsarticle}
\usepackage{color}
\usepackage{float}
\usepackage{amsmath}
\usepackage{amssymb}
\usepackage{microtype} 
\usepackage{lineno}
\begin{document}
\begin{frontmatter}

\title{Suppressing spurious oscillations and particle noise in particle-in-cell simulations}
\author[um]{Yuxi Chen}
\ead{yuxichen@umich.edu}
\cortext[cor1]{Corresponding author}
\author[bu]{Hongyang Zhou}
\author[um]{G\'abor T\'oth}
\address[um]{Department of Climate and Space Sciences and Engineering, University of Michigan, Ann Arbor, MI 48109, USA}
\address[bu]{Center for Space Physics, Boston University, Boston, MA 02215, USA}

\begin{abstract}
Particle-in-cell (PIC) simulations are essential for studying kinetic plasma processes, but they often suffer from statistical noise, especially in plasmas with fast flows. We have also found that the typical central difference scheme used in PIC codes to solve Maxwell's equations produces spurious oscillations near discontinuities, which can lead to unphysical solutions. In this work, we present numerical techniques to address these challenges within the semi-implicit PIC code FLEKS, which is based on the Gauss’s Law-satisfying Energy-Conserving Semi-Implicit Particle-in-Cell method (GL-ECSIM). First, we introduce a Lax-Friedrichs-type diffusion term with a flux limiter into the Maxwell solver to suppress unphysical oscillations near discontinuities. Second, we propose a novel approach for calculating the current density in the comoving frame, which significantly reduces particle noise in simulations with fast plasma flows. Numerical tests are presented to demonstrate the effectiveness of these methods in mitigating spurious oscillations and noise in shock and magnetic reconnection simulations.
\end{abstract}
\begin{keyword}
particle-in-cell \sep statistical noise \sep discontinuity
\end{keyword}
\end{frontmatter}

The particle-in-cell (PIC) method is a widely used computational approach for simulating kinetic plasma phenomena, including kinetic waves, magnetic reconnection, and collisionless shocks. In PIC codes, Maxwell's equations are typically solved using the Yee mesh and its associated central difference scheme, also known as the finite-difference time-domain (FDTD) method. The Yee mesh employs a staggered grid, where electric fields \(\mathbf{E}\) and magnetic fields \(\mathbf{B}\) are defined at offset locations, enabling accurate and convenient calculation of field curvatures via central differences. However, the FDTD method is inherently dispersive and prone to generating spurious oscillations near discontinuities \cite{taflove2005computational}. While such dispersive effects are often negligible in simulations without discontinuities, they can lead to significant unphysical artifacts in shock simulations.

Another major source of error in PIC simulations is statistical noise \cite{birdsall2018plasma}, which arises from the finite number of computational macro-particles. Increasing the number of particles per cell (\(n\)) reduces noise, but only with limited efficiency, as the noise scales as \(n^{-1/2}\). Employing higher-order particle shape functions can also suppress noise \cite{fiuza2011efficient, arber_contemporary_2015}, but at the cost of increased complexity in particle-mesh interpolation algorithms. For semi-implicit PIC methods, the particle-mesh interaction can be even more complicated \cite{lapenta_exactly_2017} and limits the use of high-order shape functions. Smoothing techniques, such as filtering the electric field and/or current density \cite{birdsall2018plasma,arber_contemporary_2015}, are commonly used to mitigate noise, but they introduce additional numerical diffusion and require careful tuning of the smoothing parameters \cite{vay_numerical_2011, werner2025suppressing}.

In our recent application of the semi-implicit PIC code, the Flexible Exascale Kinetic Simulator (FLEKS)\cite{chen_fleks_2023}, to shock simulations, we observed that both the dispersive nature of the Maxwell solver and statistical noise can produce unphysical results. This paper presents numerical techniques developed to address these challenges. FLEKS is based on the Gauss’s Law-satisfying Energy-Conserving Semi-Implicit Particle-in-Cell method (GL-ECSIM) \cite{lapenta_exactly_2017,Chen2019a}, which utilizes a mesh and field solver distinct from the explicit Yee scheme. Nevertheless, the numerical methods introduced here are also applicable to explicit PIC codes.

Spurious oscillations due to discretization errors are a well-known issue in numerical solutions of hyperbolic equations \cite{Hirsch:1989}. For example, the forward time centered space (FTCS) scheme, which uses central differences for spatial discretization, is unconditionally unstable for hyperbolic equations. The second-order Lax-Wendroff (LW) method is conditionally stable but still produces oscillations near discontinuities. In contrast, upwind schemes and the first-order Lax-Friedrichs (LF) method can be total variation diminishing (TVD) \cite{Harten:1972}, i.e., free of oscillations, for one-dimensional (1D) linear hyperbolic equations. To achieve both accuracy and non-oscillatory solutions, flux limiter methods have been developed, which adaptively switch between high-order and low-order schemes based on local solution smoothness \cite{vanLeer:1979}. High-order approximations are used in smooth regions, while low-order schemes are applied near discontinuities to ensure the TVD property. We refer readers to \cite{leveque2002finite} for a comprehensive review of numerical methods for hyperbolic equations. Motivated by these TVD approaches, we introduce a Lax-Friedrichs-type diffusion term, combined with a flux limiter, into the Maxwell solver to suppress spurious oscillations near discontinuities. The details of this approach are described in section~\ref{sec:discretization}.

In addition, our shock simulations revealed that fast upstream plasma flows can become unstable due to particle noise. To address this, we propose a novel method for calculating noise-reduced current densities, as detailed in section~\ref{sec:particle_noise_control}. Comprehensive numerical tests are presented in section~\ref{sec:test} to demonstrate the effectiveness of these methods in mitigating spurious oscillations and statistical noise.

\section{Discretization}
\label{sec:discretization}
\subsection{Discretization of Maxwell's equations}
\label{sec:discretization_of_maxwell}
An electromagnetic PIC code solves Ampere's law and Faraday's law: 
\begin{align}
  \frac{\partial \mathbf{E}}{\partial t} &= c \nabla \times \mathbf{B} - 4 \pi \mathbf{J}, \label{eq:ampere_law} \\
  \frac{\partial \mathbf{B}}{\partial t} &= -c \nabla \times \mathbf{E}. \label{eq:faraday_law}
\end{align}
In a PIC code, plasma is represented by macro-particles, and it is coupled to Maxwell's equations through the current density \(\mathbf{J}\).

The Yee method employs an uncollocated staggered mesh \cite{yee1966numerical},
where different components of the electric and magnetic fields are defined at distinct spatial locations. In contrast, our semi-implicit method utilizes a collocated staggered grid \cite{Lapenta:2017}, with electric fields located at the grid nodes and magnetic fields at the cell centers. In the one-dimensional (1D) case, these two grid arrangements are equivalent. To illustrate the discretization of Maxwell's equations, we consider the propagation of a 1D electromagnetic wave in vacuum, i.e., with zero current density (\(\mathbf{J} = 0\)). In both the Yee mesh and our collocated staggered mesh for the 1D case, the magnetic field $B_z$ is defined at cell centers, while the electric field $E_y$ is defined at the nodes. Throughout this paper, integer indices refer to cell centers, and half-integer indices refer to grid nodes. 

In the Yee method, both the electric and magnetic fields are staggered in time as well. Specifically, the electric field is defined at half-integer time steps ($n+1/2$), while the magnetic field is defined at integer time steps ($n$). The explicit central difference scheme is employed to advance the fields in time, providing second-order accuracy for both spatial and temporal derivatives: 
\begin{align}
  E_{y,i+1/2}^{n+1/2}  &= E_{y,i+1/2}^{n-1/2} -\frac{c\Delta t}{\Delta x}\left(B_{z,i+1}^{n} - B_{z,i}^{n} \right), \label{eq:yee-e}\\
  B_{z,i}^{n+1} &= B_{z,i}^{n} -\frac{c\Delta t}{\Delta x}\left(E_{y,i+1/2}^{n+1/2} - E_{y,i-1/2}^{n+1/2} \right). \label{eq:yee-b}
\end{align}
Due to its dispersive nature, unphysical oscillations can occur near discontinuities \cite{taflove2005computational}.

In our semi-implicit PIC code, both electric and magnetic fields are defined at integer time steps, and the following implicit $\theta$-method is used: 
\begin{linenomath} \begin{eqnarray}
  \frac{\mathbf{B}^{n+1}-\mathbf{B}^{n}}{\Delta t} &=&
        -c\nabla \times \mathbf{E}^{n+\theta}, 
  \label{eq:theta-B}\\
  \frac{\mathbf{E}^{n+1}-\mathbf{E}^{n}}{\Delta t} &=& 
  c\nabla \times \mathbf{B}^{n+\theta}  - 4\pi \mathbf{\bar{J} },
  \label{eq:theta-E}
  \end{eqnarray}\end{linenomath}
where $\theta \in [0.5, 1]$. It achieves second-order accuracy with $\theta = 0.5$. In our code, we substitute the expression for the magnetic field $\mathbf{B}^{n+\theta}=(1-\theta)\mathbf B^n+\theta \mathbf B^{n+1}$ into Ampere's law, and solve the following equations:
\begin{align}
  \mathbf{E}^{n+\theta} + (c \theta \Delta t)^2 \left[ \nabla(\nabla \cdot \mathbf{E}^{n+\theta}) - 
  \nabla ^2 \mathbf{E}^{n+\theta}\right] &=\mathbf{E}^{n} 
     + c \theta \Delta t \left(\nabla \times \mathbf{B}^n - \frac{4\pi}{c}\mathbf{\bar{J}}\right),
  \label{eq:ECSIM-E} \\
  \mathbf{B}^{n+1} &= \mathbf{B}^{n} 
      -c \theta \Delta t\nabla \times \mathbf{E}^{n+\theta}. \label{eq:ECSIM-B} 
    \end{align}      
We refer readers to \cite{chen_fleks_2023} for a detailed derivation. In the absence of plasma (i.e., in vacuum), the current density vanishes and the divergence of the electric field is zero. Under these conditions, the one-dimensional discretization of the electric field reduces to
\begin{equation}
  E_{y,i+1/2}^{n+\theta} - \frac{(c \theta \Delta t)^2}{(\Delta x)^2} \left( E_{y,i+3/2}^{n+\theta} - 2E_{y,i+1/2}^{n+\theta} + E_{y,i-1/2}^{n+\theta} \right) = E_{y,i+1/2}^{n} - \frac{c \theta \Delta t}{\Delta x}\left(B_{z,i+1}^{n} - B_{z,i}^{n} \right). \label{eq:e-1d}
\end{equation}
Without the left-hand side (LHS) second-order spatial derivative, Eq.~\ref{eq:e-1d} reduces to a form that closely resembles the explicit update in Eq.~\ref{eq:yee-e}, with the distinction that both the electric and magnetic fields are defined at the same time step.
 
It is interesting to note that Eq.~\ref{eq:e-1d} is similar to the second-order Lax-Wendroff (LW) method. To demonstrate this correspondence, we first derive the equation for $E_y^{n+1}$. Recall that $\mathbf E^{n+\theta}$ is defined as
 \begin{equation}
\mathbf E^{n+\theta} = (1-\theta) \mathbf E^n + \theta \mathbf E^{n+1}.
 \end{equation}
Eq. \ref{eq:e-1d} can be rewritten in the following form by assuming $\theta=0.5$:  
\begin{equation}
  E_{y,i+1/2}^{n+1} - \frac{c^2\Delta t^2}{2\Delta x^2} \left( E_{y,i+3/2}^{n+1} - 2E_{y,i+1/2}^{n+1} + E_{y,i-1/2}^{n+1} \right) = E_{y,i+1/2}^{n} - \frac{c\Delta t}{\Delta x}\left(B_{z,i+1}^{n} - B_{z,i}^{n} \right). \label{eq:e-1d-lw}
\end{equation}
On the other hand, for the one-dimensional linear advection equation $\frac{\partial u}{\partial t} + c\frac{\partial u}{\partial x} = 0$, the Lax-Wendroff method is given by
\begin{equation}
  u_{i+1/2}^{n+1} = u_{i+1/2}^n - \frac{c\Delta t}{2\Delta x}\left(u^{n}_{i+3/2} - u^{n}_{i-1/2} \right) + \frac{c^2\Delta t^2}{2\Delta x^2}\left(u^{n}_{i+3/2} - 2u^{n}_{i+1/2} + u^{n}_{i-1/2} \right). \label{eq:l-w}
\end{equation}
For easier comparison with Eq.~\ref{eq:e-1d-lw}, here $u$ is defined on the half-integer grid. The second-order spatial derivative on the right-hand side (RHS) of Eq.~\ref{eq:l-w} is treated explicitly, whereas the corresponding term on the left-hand side (LHS) of Eq.~\ref{eq:e-1d-lw} is treated implicitly. However, both terms share the same coefficient. It is well known that the Lax-Wendroff method exhibits dispersive oscillations near discontinuities. The implicit scheme in Eq.~\ref{eq:e-1d-lw} shows similar oscillatory behavior, as will be demonstrated in the numerical tests presented in section~\ref{sec:test}.

Once the electric field is obtained from the implicit solver, the magnetic field is advanced in time using an explicit update: 
\begin{equation}
  B_{z,i}^{n+1} = B_{z,i}^{n} -\frac{c\Delta t}{\Delta x}\left(E_{y,i+1/2}^{n+\theta} - E_{y,i-1/2}^{n+\theta} \right). \label{eq:1d-b}
\end{equation}
When $\theta=0.5$, the expression above is the same as the Yee update of Eq. \ref{eq:yee-b}.

\subsection{Suppressing spurious oscillations}
\label{sec:suppress_oscillation}

For a one-dimensional (1D) linear hyperbolic equation, \(\frac{\partial u}{\partial t} + c\frac{\partial u}{\partial x} = 0\), the forward time-centered space (FTCS) scheme is:
\begin{equation}
  u_{i}^{n+1} = u_{i}^n - \frac{\Delta t}{\Delta x}\left(f_{c,i+1/2}^n - f_{c,i-1/2}^n\right),
\end{equation}
where \(f_{c,i+1/2}^n = \frac{c}{2}(u_{i+1}^n + u_i^n)\). This scheme is unconditionally unstable for hyperbolic problems. The Lax-Friedrichs (LF) method introduces a numerical diffusion term to stabilize the scheme. The LF flux is defined as:
\begin{equation}
  f_{\text{LF},i+1/2}^n = f_{c,i+1/2}^n - \frac{c}{2}\left(u_{i+1}^n - u_i^n\right).
\end{equation}
The Lax-Friedrichs method is total variation diminishing (TVD) and free of spurious oscillations for 1D linear equations. To reduce numerical diffusion, the diffusion coefficient $c$ can be replaced by the local maximum characteristic speed \(a_{i+1/2}\), resulting in the local Lax-Friedrichs (LLF) or Rusanov scheme:
\begin{equation}
  f_{\text{LLF},i+1/2}^n = f_{c,i+1/2}^n - \frac{a_{i+1/2}}{2}\left(u_{i+1}^n - u_i^n\right).
\end{equation}
Motivated by the Lax-Friedrichs approach, we introduce a diffusion term to the right-hand side (RHS) of the magnetic field update in Eq.~\ref{eq:1d-b} to suppress spurious oscillations:
\begin{equation}
  B_{z,i}^{n+1} = B_{z,i}^{n} -\frac{\Delta t}{\Delta x}\left(
    f_{B, i+1/2}^n - f_{B,i-1/2}^n 
    \right). \label{eq:1d-b-diss}  
\end{equation}
Here, the numerical flux is defined as
\begin{equation}
  f_{B,i+1/2}^n = cE_{y,i+1/2}^{n+\theta} - \frac{(1-\phi_{B,i+1/2})a_{B,i+1/2}}{2}\left(B_{z,i+1}^n - B_{z,i}^n\right). \label{eq:flux-b}
\end{equation}
In this expression, \(\phi_{B,i+1/2}\) is a flux limiter function (discussed below), and \(a_{B,i+1/2}\) is the local diffusion speed. When 
\(\phi_{B,i+1/2}=0\), the first order scheme is employed,
while \(\phi_{B,i+1/2}=1\) corresponds to the original second order scheme.

Similarly, an implicit diffusion term can be incorporated into the left-hand side (LHS) of the electric field solver in Eq.~\ref{eq:e-1d}:
\begin{align}
  E_{y,i+1/2}^{n+\theta} - \frac{(c \theta \Delta t)^2}{\Delta x^2} \left( E_{y,i+3/2}^{n+\theta} - 2E_{y,i+1/2}^{n+\theta} + E_{y,i-1/2}^{n+\theta} \right) +
  \frac{\theta \Delta t}{\Delta x}\left( d_{i+1}^{n+\theta} - d_{i}^{n+\theta} \right)  = \nonumber \\ 
  E_{y,i+1/2}^{n} - \frac{c \theta \Delta t}{\Delta x}\left(B_{z,i+1}^{n} - B_{z,i}^{n} \right)
\end{align}  
where \(d_{i}^{n+\theta}\) is the numerical flux corresponding to the diffusion term:
\begin{equation}
  d_{i}^{n+\theta} = \frac{(1-\phi_{E,i}) a_i}{2} \left(E_{y,i+1/2}^{n+\theta} - E_{y,i-1/2}^{n+\theta}\right).  \label{eq:D-E} 
\end{equation}
Here, \(\phi_{E,i+1}\) and \(a_{E, i+1}\) are the flux limiter and local diffusion speed, respectively.

We use van Leer's generalized monotonized central (MC) limiter function \cite{vanLeer:1979} with a tunable parameter \(\beta\) for both the electric (\(\phi_E\) in Eq.~\ref{eq:D-E}) and magnetic (\(\phi_B\) in Eq.~\ref{eq:flux-b}) fields:
\begin{equation}
  \phi(r,\beta) = \max\left(0,\min\left(\beta r,\frac{1+r}{2},\beta\right)\right),\quad\beta\in\left[1,2\right].
\end{equation}
where \(r\) is a local smoothness indicator, defined for the magnetic field limiter \(\phi_{B,i+1/2}(r)\) as
\begin{equation}
 r = \frac{B_{z,i}^n - B_{z,i-1}^n}{B_{z,i+1}^n - B_{z,i}^n},
 \end{equation}
and for the electric field limiter \(\phi_{E,i}(r)\) as
 \begin{equation}
 r = \frac{E_{y,i-1/2}^{n+\theta} - E_{y,i-3/2}^{n+\theta}}{E_{y,i+1/2}^{n+\theta} - E_{y,i-1/2}^{n+\theta}}.  
 \end{equation}
A larger \(\beta\) value results in a less diffusive limiter. \(\beta=1\) corresponds to the minmod limiter, while \(\beta=0\) results in the unlimited first order scheme.

For an explicit scheme, the diffusion speed \(a\) should be set to the local maximum characteristic speed, which is the speed of light in vacuum. Our numerical experiments show that using $a=c$ leads to diffusive solutions. After some trials, we found that using the local plasma bulk speed for $a=u$ provides an optimal balance between accuracy and stability for shock and magnetic reconnection simulations.
This choice is also motivated by the observation that the time
step is based on the particle velocities and not on the
electromagnetic wave speed.

\subsection{$\nabla \cdot \mathbf{B}$ correction}
The addition of the diffusion term in the magnetic field update (Eq.~\ref{eq:flux-b}) breaks the solenoidal constraint, $\nabla \cdot \mathbf{B} = 0$. To mitigate the growth of magnetic field divergence, we use the hyperbolic/parabolic divergence cleaning method \cite{Dedner:2001} to control the growth of the magnetic field divergence errors. Specifically, we solve the following mixed hyperbolic-parabolic equation for $\varphi$:
\begin{equation}
  \frac{\partial \varphi}{\partial t} + c_h^2 \nabla \cdot \mathbf{B} = -\frac{c_h^2}{c_p^2} \varphi,
\end{equation}
where $c_h$ is the hyperbolic cleaning speed and $c_p$ is the parabolic damping parameter. $\varphi$ is then used to correct the magnetic field as follows:
\begin{equation}
  \frac{\mathbf{B}^{n+1}-\mathbf{B}^{n}}{\Delta t} =
        -c\nabla \times \mathbf{E}^{n+\theta}  - \nabla \varphi.
\end{equation}
 In our simulations, we set $c_h = 0.8\,\Delta x/\Delta t$ and choose $\Delta t\, c_h^2/c_p^2 = 0.1$. The scalar function $\varphi$ is defined at the cell centers similarly to the magnetic field. The numerical scheme for solving the $\varphi$ equation follows the same approach as described in \cite{toth_adaptive_2012}. Further details on the hyperbolic/parabolic divergence cleaning method can be found in \cite{Dedner:2001}.

\subsection{From smoothing fields to the Lax-Friedrichs diffusion}
In this section, we discuss the rationale for adopting the Lax-Friedrichs diffusion term to suppress spurious oscillations. While the original GL-ECSIM method performs well for magnetic reconnection simulations \cite{chen_global_2017,Chen2020,Chen2019}, it yields unphysical results when applied to shock problems. Our initial approach involved smoothing the electric field components (here denoted by the generic variable $v$) using a digital filter:
\begin{equation}
v^*_{i+1/2} = \phi v_{i+1/2} + \frac{1-\phi}{2}(v_{i+1/2-d} + v_{i+1/2+d}), \label{eq:filter}  
\end{equation}
with default parameters $\phi=0.5$ and $d=1$. Applying this filter for multiple passes improved the stability of the simulations, allowing them to run longer before producing unphysical waves. However, persistent oscillations still developed over time in many cases. We also experimented with Vay's filter \cite{vay_mesh_2002}, which combines different strides $d$ to enhance smoothing. Although Vay's filter is more effective than the standard binomial filter, it still does not completely suppress numerical oscillations.

Further investigation revealed that applying the digital filter even once to the magnetic field, in addition to the electric field, is highly effective at suppressing oscillations. However, this approach introduces excessive numerical diffusion, significantly smearing sharp features. A key limitation of the digital filter is its independence from the time step: the same amount of smoothing is applied regardless of $\Delta t$, which can lead to large diffusion, especially for small time steps. To address this, we considered a time-step-dependent filter coefficient, $\phi \sim \frac{\Delta t c}{\Delta x}$, which naturally leads to a Lax-Friedrichs-type diffusion term. Replacing the global characteristic speed $c$ with a local characteristic speed $a$ further reduces diffusion, resulting in the local Lax-Friedrichs (LLF) diffusion term. Finally, introducing the flux limiter, further reduces the diffusion and preserves the second-order accuracy of the field solver.

After successfully implementing the second order local Lax-Friedrichs diffusion term in the magnetic field solver, we extended this approach to the electric field solver by replacing digital filter smoothing with the implicit Lax-Friedrichs diffusion term in Eq.~\ref{eq:D-E}. This modification yields solutions that are less diffusive and more accurate, as demonstrated by the light wave test in section~\ref{sec:test}. Consequently, we adopted the second-order local Lax-Friedrichs diffusion term for both the electric and magnetic field solvers as our final approach to suppressing spurious oscillations while preserving sharp discontinuities.
  
\section{Particle noise control}
\label{sec:particle_noise_control}
\subsection{Calculating current in the comoving frame}
Particle noise is a major source of numerical error in PIC simulations. In this section, we introduce a method to reduce noise in simulations that contains fast plasma flows, such as the solar wind, where the thermal speed is much smaller than the bulk flow speed.

In electromagnetic PIC codes, particles are coupled to the electromagnetic fields through the current density \(\mathbf{J}_g\) at each grid location \(\mathbf{x}_g\):
\begin{equation}
  \mathbf{J}_g = \frac{1}{\Delta V} \sum_{s,p} q_{s,p}\mathbf{v}_{s,p} W(\mathbf{x}_{s,p} - \mathbf{x}_g), 
\end{equation} 
where $s$ denotes the particle species, $g$ and $p$ are the grid point and particle indices, $q_{s,p}$ and $\mathbf{v}_{s,p}$ are the macro-particle charge and velocity, $W(\mathbf{x}_{s,p} - \mathbf{x}_g)$ is the interpolation function, and $\Delta V$ is the grid cell volume. In fast flows, statistical noise in the current density can become substantial. To illustrate this, consider a uniform fast flow with constant bulk velocity $\mathbf{u}$:
\begin{eqnarray}
  \mathbf{J}_g &=& \frac{1}{\Delta V} \sum_{s,p} q_{s,p}\mathbf{v}_{s,p} W(\mathbf{x}_{s,p} - \mathbf{x}_g), \nonumber \\
  &=& \frac{1}{\Delta V} \sum_{s,p} q_{s,p}(\mathbf{v}_{s,p} - \mathbf{u}) W(\mathbf{x}_{s,p} - \mathbf{x}_g) + \frac{1}{\Delta V} \mathbf{u} \sum_{s,p} q_{s,p} W(\mathbf{x}_{s,p} - \mathbf{x}_g) \nonumber \\
  &=& \frac{1}{\Delta V} \sum_s q_{s,g} (\mathbf{u}_{s,g} - \mathbf{u}) + \frac{1}{\Delta V} \left( \sum_s q_{s,g} \right) \mathbf{u}, \label{eq:jn}
\end{eqnarray}
where
\begin{eqnarray}
 q_{s,g} &=& \sum_{p} q_{s,p} W(\mathbf{x}_{s,p} - \mathbf{x}_g), \nonumber \\
  \mathbf{u}_{s,g} &=& \frac{1}{q_{s,g}} \sum_{p} q_{s,p}\mathbf{v}_{s,p} W(\mathbf{x}_{s,p} - \mathbf{x}_g). \nonumber 
\end{eqnarray}
Here, $q_{s,g}$ and $\mathbf{u}_{s,g}$ are the charge and average velocity of species $s$ at grid point $\mathbf{x}_g$, respectively. The first term in Eq.~\ref{eq:jn} represents the current due to the velocity difference between ions and electrons, while the second term corresponds to the current associated with the transport of net charge. In a physically uniform plasma, both terms should vanish. However, in PIC simulations, statistical fluctuations in particle positions and velocities lead to nonzero values for both terms. Since $|\mathbf{u}| \gg |\mathbf{v}_{s,p}-\mathbf{u}|$ in fast flows, the noise contribution from the second term can dominate, resulting in large unphysical current fluctuations.

If we perform simulations in a coordinate system moving with the bulk velocity $\mathbf{u}$, the second term in Eq.~\ref{eq:jn} vanishes, which is desirable for uniform flows. Motivated by this, we propose a method to reduce current noise by retaining only the current in the comoving frame, i.e., by neglecting the second term in Eq.~\ref{eq:jn}. Physically, this approach assumes that the current associated with the motion of the net charge (second term in Eq.~\ref{eq:jn}) is negligible compared to the current arising from the velocity difference between ions and electrons (first term in Eq.~\ref{eq:jn}). It is important to note that net charge can still develop due to current divergence. While this assumption is not universally valid, it offers advantages for space plasma magnetic reconnection and shock simulations for the following reasons:
\begin{itemize}
  \item Charge separation occurs at the Debye length scale, whereas reconnection and kinetic shock processes typically occur at the electron or ion inertial length scales, which are much larger. In cold plasmas with low thermal speeds, the Debye length can be several orders of magnitude smaller than the inertial lengths, making the net charge contribution to the current negligible at the scales of interest.
  \item In practice, reconnection and shock simulations do not resolve the physical Debye length, nor do they accurately capture the net charge. The net charge density produced by statistical noise is on the order of $n^{-1/2} q_{g,e}/\Delta V$, where $n$ is the number of particles per cell. For a typical value of $n=100$, the noise-induced net charge density is about $10\%$ of the electron charge density. Since the simulation grid size is usually much larger than the Debye length, the noise-induced net charge is much larger than the physical net charge. Therefore, neglecting the current associated with the motion of the net charge is reasonable in practice.
\end{itemize}
Numerical tests in section~\ref{sec:test} demonstrate that this assumption stabilizes simulations without introducing significant errors.

It is straightforward to apply this assumption to an explicit PIC code, as the current density is directly computed from known particle positions and velocities. In general, the comoving frame velocity $\mathbf{u}(\mathbf{x})$ varies spatially, and the current density at each grid point is given by
\begin{eqnarray}  
  \mathbf{J}_g &=& \frac{1}{\Delta V} \sum_{s,p} q_{s,p} \left[\mathbf{v}_{s,p} - \mathbf{u}(\mathbf{x}_{s,p})\right] W(\mathbf{x}_{s,p} - \mathbf{x}_g) + \frac{1}{\Delta V} \sum_{s,p} q_{s,p} \mathbf{u}(\mathbf{x}_{s,p}) W(\mathbf{x}_{s,p} - \mathbf{x}_g). \nonumber
\end{eqnarray}
By omitting the second term, which represents the current associated with the transport of net charge, the current in the comoving frame is obtained. The comoving frame velocity $\mathbf{u}(\mathbf{x})$ is determined by smoothing the plasma velocity using the filter defined in Eq.~\ref{eq:filter}. Typically, we applied five times of smoothing with $\phi=0.5$ and alternating between $d=1$ and $d=2$ to ensure that the velocity is smooth.

For semi-implicit PIC codes, calculating the comoving frame current is more involved because the current depends on particle velocities that are not known yet. In our semi-implicit PIC code, Eq.~\ref{eq:ECSIM-E} is solved, where $\mathbf{\bar{J}}$ is the predicted current at the $n+\frac{1}{2}$ time step:
\begin{eqnarray}
  \mathbf{\bar{J}}_g &=& \frac{1}{\Delta V} \sum_{s,p} q_{s,p} \mathbf{\bar{v}_{s,p}}W(\mathbf{x}_{s,p} - \mathbf{x}_g),
\end{eqnarray}
with
\begin{eqnarray}
  \mathbf{\bar{v}}_{s,p} &=& \mathbf{\hat{v}}_{s,p} + \beta_s \mathbf{\hat{E}}_{s,p}, \nonumber \\
  \mathbf{\hat{v}}_{s,p} &=& \alpha^n_p \cdot \mathbf{v}_{s,p}^n, \nonumber \\
  \mathbf{\hat{E}}_{s,p} &=& \alpha^n_p \cdot \mathbf{E}_{s,p}^{n+\theta}. \nonumber
\end{eqnarray}
Here, $\beta_s = (q_s/m_s)(\Delta t/2)$ and $\alpha^n_p$ is a rotation matrix:
\begin{equation}
\alpha^n_p = \frac{1}{1+(\beta_s B^n_p)^2} \left(\mathbb{I} - \beta_s \mathbb{I} \times \mathbf{B}^n_p + \beta_s^2 \mathbf{B}^n_p\mathbf{B}^n_p  \right),
\end{equation}
where $\mathbb{I}$ is the identity dyadic. For further details on the derivation, we refer the reader to \cite{lapenta_exactly_2017}.

Assuming the plasma bulk velocity is $\mathbf{u}(\mathbf{x})$, in the comoving frame the particle velocity is $\mathbf{v}_{s,p,\text{C}} = \mathbf{v}_{s,p} - \mathbf{u}(\mathbf{x}_{s,p})$, and the electric field is $\mathbf{E}_\text{C} = \mathbf{E} - \mathbf{E}_\text{M}$, where 
$\mathbf{E}_\text{M} = -\mathbf{u} \times \mathbf{B}$ is the motional electric field. Thus, the current $\mathbf{\bar{J}}$ can be rewritten as:
\begin{eqnarray}
  \mathbf{\bar{J}}_g = \frac{1}{\Delta V} \sum_{s,p} q_{s,p} \alpha_p^n \left[\mathbf{v}_{s,p,\text{C}}^n + \beta_s\mathbf{E}^{n+\theta}_{s,p,\text{C}} + \mathbf{u}(\mathbf{x}_{s,p}) + \beta_s\mathbf{E}_\text{M} (\mathbf{x}_{s,p}) \right]W(\mathbf{x}_{s,p} - \mathbf{x}_g).
\end{eqnarray}
After some calculations, it can be shown that $\alpha_p^n\left[ \mathbf{u}(\mathbf{x}_{s,p}) + \beta_s\mathbf{E}_\text{M} (\mathbf{x}_{s,p}) \right] = \mathbf{u}(\mathbf{x}_{s,p})$, so $\mathbf{\bar{J}}_g$ simplifies to
\begin{eqnarray}
  \mathbf{\bar{J}}_g =&& \frac{1}{\Delta V} \sum_{s,p} q_{s,p} \alpha_p^n \left[\mathbf{v}_{s,p,\text{C}}^n + \beta_s\mathbf{E}^{n+\theta}_{s,p,\text{C}}  \right]W(\mathbf{x}_{s,p} - \mathbf{x}_g) + \nonumber \\
   &&\frac{1}{\Delta V} \sum_{s,p} q_{s,p} \mathbf{u}(\mathbf{x}_{s,p})  W(\mathbf{x}_{s,p} - \mathbf{x}_g).
\end{eqnarray}
By neglecting the second term, which corresponds to the current generated by the motion of the net charge, we obtain the comoving frame current density. This can be written as:
\begin{equation}
  \mathbf{\bar{J}}_g = \mathbf{\hat{J}_{g,\text{C}}} + \frac{1}{\Delta V} \sum_{s,g^\prime}\beta_s M_{s,gg^\prime} \mathbf{E}^{n+\theta}_{g^\prime, \text{C}},
\end{equation}
where
\begin{eqnarray}
\mathbf{\hat{J}_{g,\text{C}}} = \frac{1}{\Delta V} \sum_{s,p} q_{s,p} \alpha_p^n \mathbf{v}_{s,p,\text{C}}^n W(\mathbf{x}_{s,p} - \mathbf{x}_g),
\end{eqnarray}
and $M_{s,gg^\prime} = \sum_{p} q_{s,p} \alpha^n_p W_{pg^\prime} W_{pg} $ is the mass matrix \cite{lapenta_particle_2012}. Note that Eq.~\ref{eq:ECSIM-E} solves for $\mathbf{E}^{n+\theta}$ rather than $\mathbf{E}^{n+\theta}_\text{C}$. Since $\mathbf{E}_\text{C} = \mathbf{E} - \mathbf{E}_\text{M}$, the electric field solver Eq.~\ref{eq:ECSIM-E} becomes: 
\begin{eqnarray}
  \mathbf{E}^{n+\theta} + (c \theta \Delta t)^2 \left[ \nabla(\nabla \cdot \mathbf{E}^{n+\theta}) - 
  \nabla ^2 \mathbf{E}^{n+\theta}\right] + \frac{4\pi}{c\Delta V} \sum_{s,g^\prime}\beta_s M_{s,gg^\prime} \mathbf{E}^{n+\theta}_{g^\prime} = \nonumber \\
  \mathbf{E}^{n} 
     + c \theta \Delta t \left[\nabla \times \mathbf{B}^n - \frac{4\pi}{c} \left(\mathbf{\hat{J}_{g,\text{C}}} - \frac{1}{\Delta V} \sum_{s,g^\prime}\beta_s M_{s,gg^\prime} \mathbf{E}_{g^\prime, \text{M}}  \right)  \right].
\end{eqnarray}

\subsection{Smoothing the current}
Calculating the current in the comoving frame effectively removes the artificial current associated with the transport of net charge, which arises from statistical particle noise. However, this approach does not mitigate the noise present in the first term of Eq.~\ref{eq:jn}, which originates from fluctuations in the relative velocities of ions and electrons. To further suppress this residual noise and enhance numerical stability, we apply digital filter smoothing to the comoving frame current density $\mathbf{\hat{J}_{g,C}}$. Empirically, performing one to three passes of smoothing is sufficient to significantly improve the stability and accuracy of the simulations, as demonstrated in section~\ref{sec:test}.

\section{Numerical tests}
\label{sec:test}

\subsection{Advection of discontinuities}
\begin{figure}[H]
  \centering
  \includegraphics[width=1.0\textwidth]{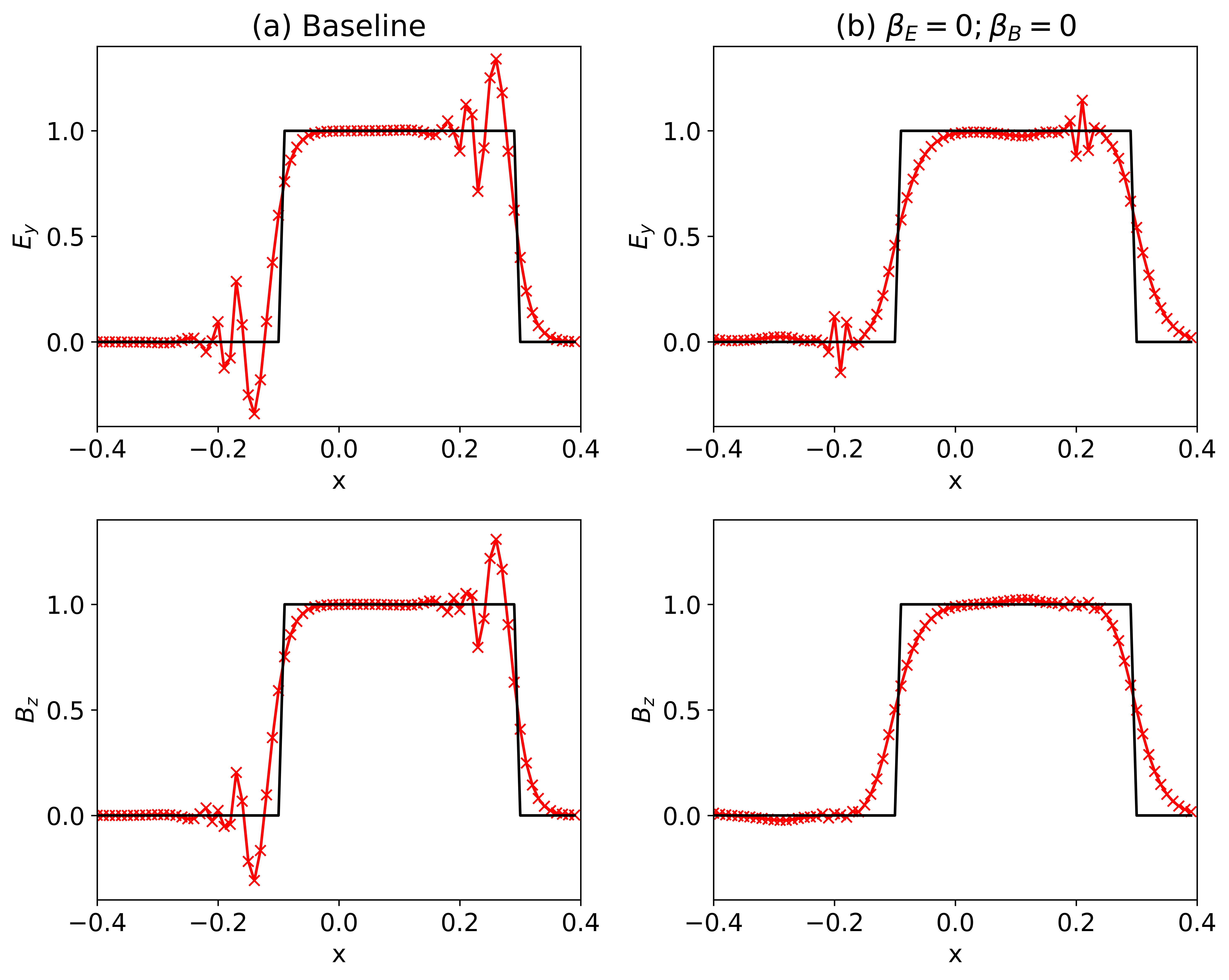}
  \caption{Light wave simulations with CFL=2 at $t=0.1$.}
  \label{fig:tophat_cfl2}
\end{figure}

This section quantitatively examines the effectiveness of the Lax-Friedrichs diffusion in controlling spurious oscillations near discontinuities. To isolate the effects of numerical diffusion from particle noise, we consider the propagation of electromagnetic waves with discontinuities in a vacuum. The computational domain is $[-0.4, 0.4]$ with a uniform grid spacing of $0.01$. The initial condition is a top-hat profile:
\begin{eqnarray}
  E_y(x) = \left\{
    \begin{array}{ll}
      1, & |x| < 0.2, \\
      0, & |x| \geq 0.2.
    \end{array}
  \right. \\
  B_z(x) = \left\{
    \begin{array}{ll}
      1, & |x| < 0.2, \\
      0, & |x| \geq 0.2.
    \end{array}
  \right.  
\end{eqnarray}
The CFL number is defined using the light speed $c=1$. Unless otherwise noted, the results correspond to a CFL number of 0.4; a higher CFL number of 2 is considered in Figure~\ref{fig:tophat_cfl2}.

\begin{figure}[H]
  \centering
  \includegraphics[width=1.0\textwidth]{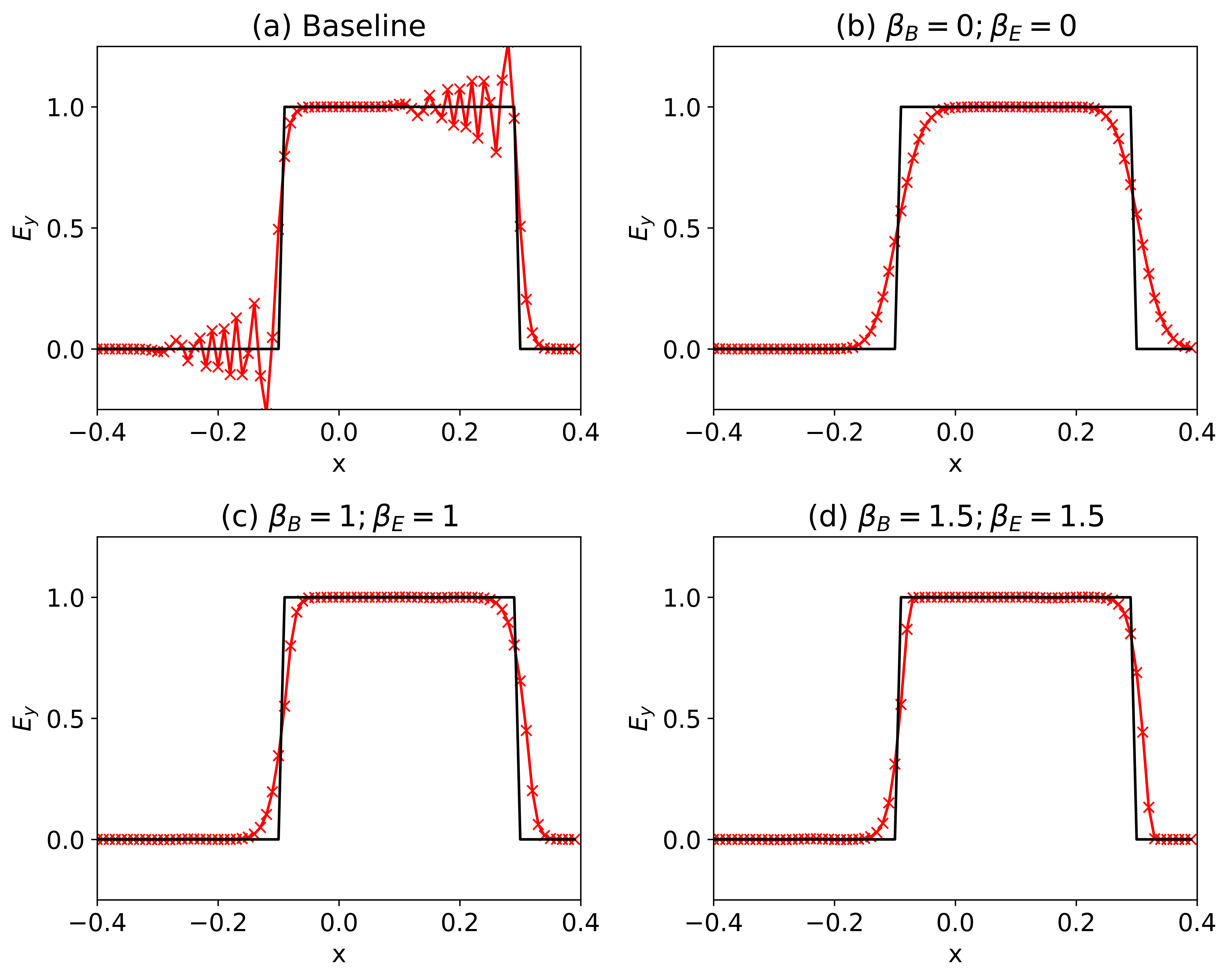}
   \caption{Light wave with discontinuities simulation results at $t=0.1$. Panel (a) is the baseline simulation without the Lax-Friedrichs diffusion. Panels (b)-(d) are the simulations with the Lax-Friedrichs diffusion and flux limiter. The flux limiter is inactive with $\beta=0$.}
  \label{fig:tophat_limiter}
\end{figure}

Figure~\ref{fig:tophat_limiter} presents the numerical solutions after 25 time steps ($t=0.1$). The reference simulation (panel a), which does not include the Lax-Friedrichs diffusion, exhibits pronounced nonphysical oscillations at the discontinuities. Introducing the Lax-Friedrichs diffusion (panels b-d) completely suppresses these oscillations. In panel b, the flux limiter is disabled, resulting in a more diffusive solution compared to panels c and d, where the flux limiter is active. Increasing the flux limiter parameter $\beta$ yields slightly sharper discontinuities, but the overall difference is minor for this test case. These results demonstrate that the combination of Lax-Friedrichs diffusion and a flux limiter effectively eliminates spurious oscillations while preserving the integrity of sharp gradients.

\begin{figure}[H]
  \centering
  \includegraphics[width=1.0\textwidth]{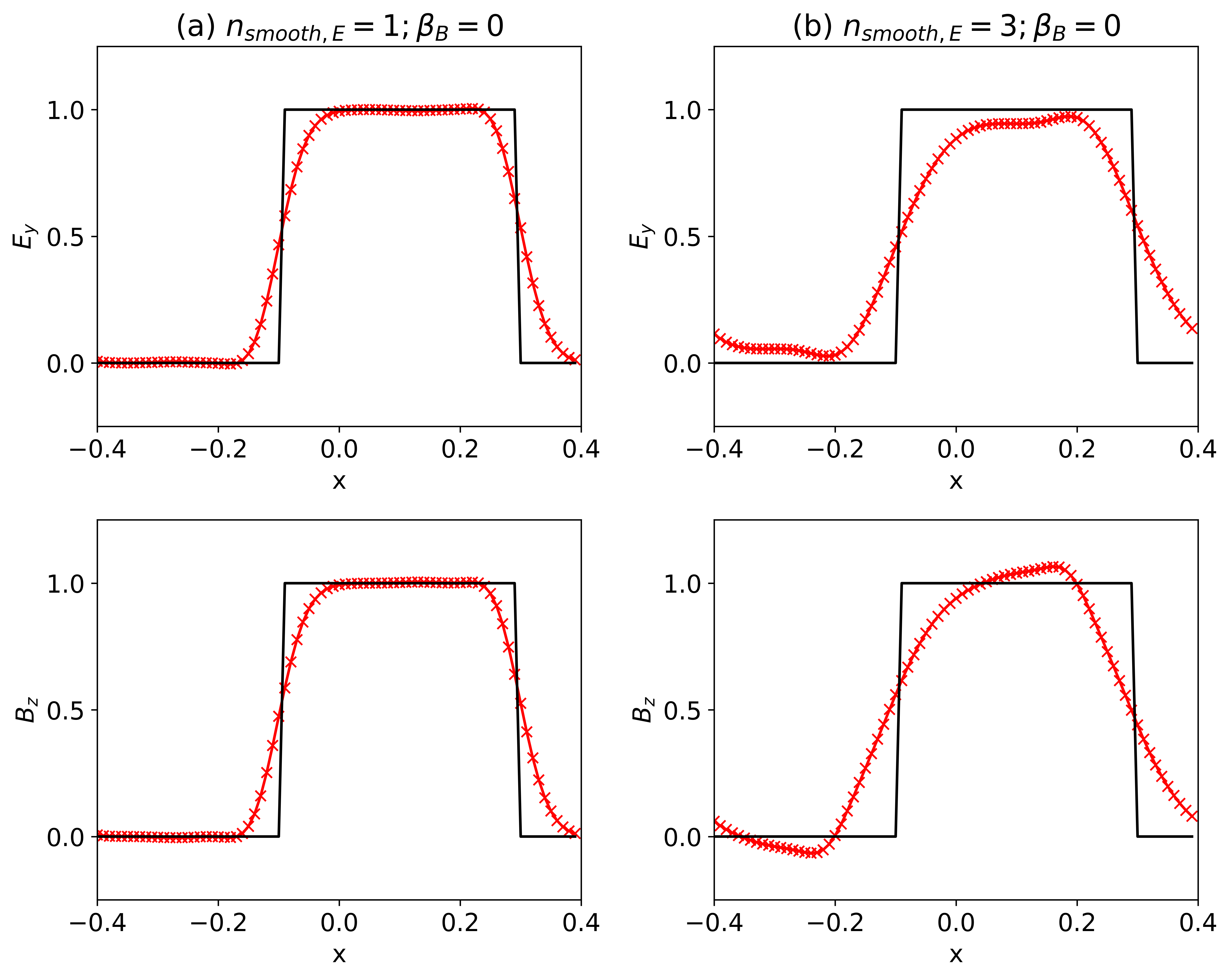}
   \caption{Light wave simulations with smoothing the electric field using a digital filter for $n_{\text{smooth},E}$ passes.}
  \label{fig:tophat_smooth}
\end{figure}

For comparison, Figure~\ref{fig:tophat_smooth} shows the results at $t=0.4$ when the electric field is smoothed using a digital filter instead of applying Lax-Friedrichs diffusion. The Lax-Friedrichs diffusion is still applied to the magnetic field. Panel a shows the results after a single smoothing pass with $\phi=0.5$ and $d=1$, while panel b shows the result after three smoothing passes with $\phi=0.5$ and alternating $d=1$ and $d=2$. The single-pass smoothing produces results comparable to the unlimited Lax-Friedrichs diffusion in Figure~\ref{fig:tophat_limiter}b. However, applying the filter three times introduces excessive numerical diffusion, significantly smearing the discontinuity. Moreover, it produces overshoot and undershoot near the discontinuity, which can generate unphysical waves in plasma simulations. Comparing Figures~\ref{fig:tophat_limiter} and \ref{fig:tophat_smooth} indicates that Lax-Friedrichs diffusion with a flux limiter is more accurate and robust than digital filter smoothing for suppressing oscillations at discontinuities.

In typical semi-implicit PIC simulations, the CFL number is determined by the plasma characteristic speed, which is much less than the speed of light. As a result, the CFL number for electromagnetic waves can be larger than 1. The implicit electric field solver remains stable and accurate for smooth solutions. Figure~\ref{fig:tophat_cfl2} shows the results for discontinuity propagation with CFL=2 at $t=0.1$, comparing cases without (panel a) and with (panel b) Lax-Friedrichs diffusion. While the Lax-Friedrichs diffusion does not entirely eliminate oscillations, it still significantly reduces oscillation amplitude. 

In practical plasma simulations, discontinuities such as shock fronts propagate at speeds that are comparable to the plasma speed. Their propagations are well resolved by the time step, and we do not need to worry about the oscillations are not fully suppressed for CFL numbers larger than 1.

\subsection{Uniform fast flow}
\begin{figure}
  \centering
  \includegraphics[width=1.0\textwidth]{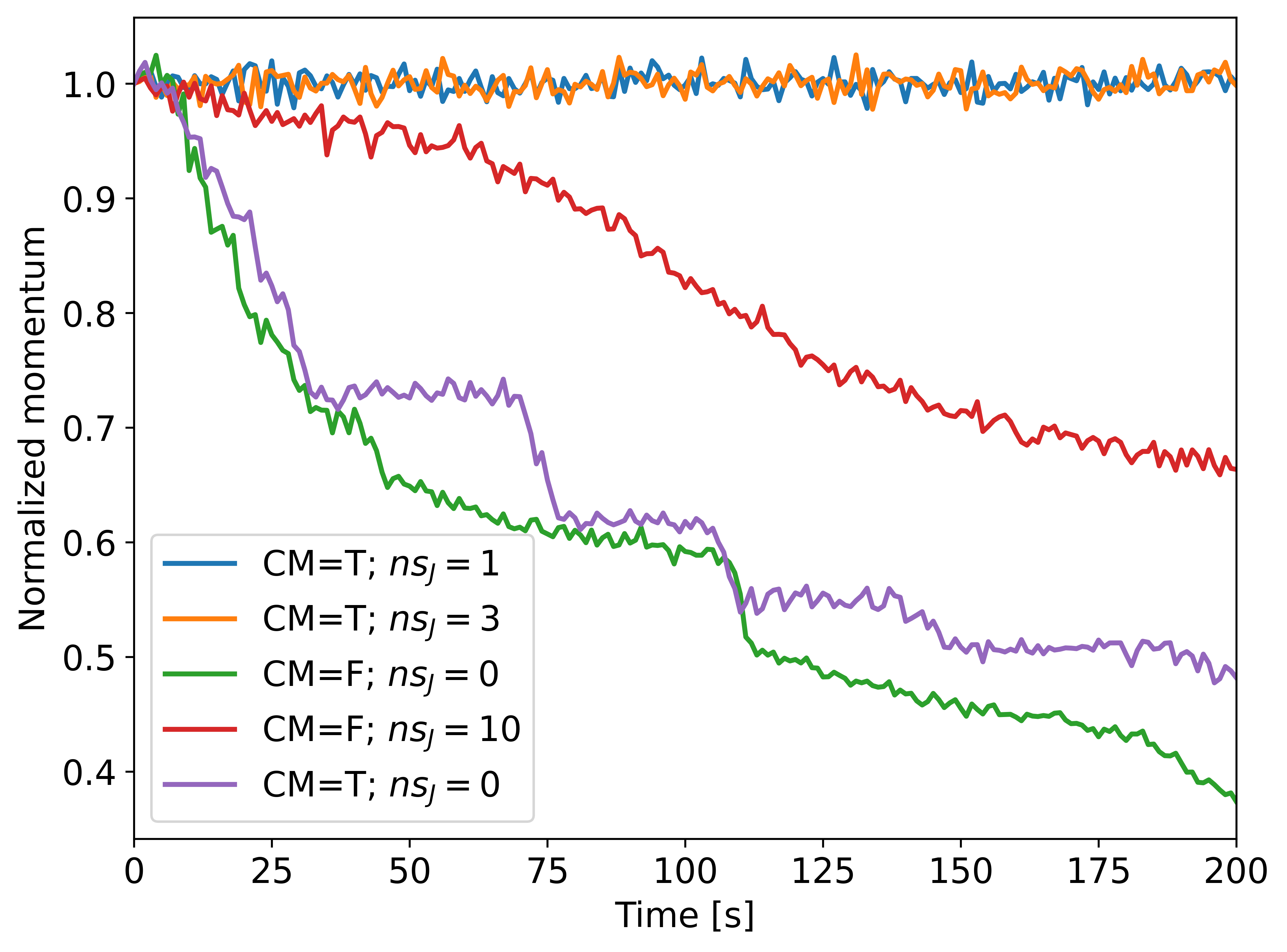}
  \caption{Time evolution of the normalized total particle momentum in uniform fast flow simulations. The legend ``CM=T'' indicates that the current is calculated in the comoving frame, and $ns_j$ is the number of digital filter smoothing passes applied to the current density.}
  \label{fig:fast_flow}
\end{figure}

This 1D test evaluates the stability of the algorithm for a uniform fast plasma flow. The initial conditions are representative of the typical solar wind at 1 AU: density $n=5\,\mathrm{cm}^{-3}$, velocity $\mathbf{u} = [400,0,0]\,\mathrm{km/s}$, magnetic field $\mathbf{B} = [0,0,5]\,\mathrm{nT}$, ion temperature $T_i = 72,000\,\mathrm{K}$, and electron temperature $T_e = 2T_i$. The ion-electron mass ratio is set to $m_i/m_e = 100$. The corresponding ion thermal speed is $V_s = 32\,\mathrm{km/s}$, Alfv\'en speed is $V_\text{A} = 49\,\mathrm{km/s}$, Alfv\'enic Mach number is $M_\text{A} = 9.2$, and the ion inertial length is $d_i = 102\,\mathrm{km}$. The simulation domain size is $160\,\mathrm{km}$, with a cell size of $8\,\mathrm{km}$. A CFL number of 0.2 is used, corresponding to a time step of approximately $0.0038\,\mathrm{s}$. The normalized total particle momentum variations from different simulations are presented in Figure~\ref{fig:fast_flow}. Since the initial conditions are spatially uniform and unperturbed, the total particle momentum should be conserved throughout the simulation.

All simulations in Figure~\ref{fig:fast_flow} employ the Lax-Friedrichs diffusion with a flux limiter parameter $\beta = 1$. The effects of calculating the current in the comoving frame (CM=T) and applying digital filter smoothing to the current are investigated. When the current is not calculated in the comoving frame (CM=F), the total particle momentum is not conserved, regardless of the number of smoothing passes ($ns_j$) applied to the current. This is because statistical noise in the current generates unphysical electromagnetic fluctuations, which in turn transfer momentum from the particles to the fields. Furthermore, the total momentum, including both particle and electromagnetic field contributions, is also not conserved due to numerical errors in the solver and the current smoothing procedure. In contrast, when the current is calculated in the comoving frame (CM=T), even a single pass of digital filter smoothing ($ns_j=1$) is sufficient to maintain excellent conservation of total particle momentum and ensure numerical stability.

This test demonstrates that calculating the current in the comoving frame is highly effective for stabilizing fast plasma flows, and that moderate smoothing of the current further enhances numerical stability.

\subsection{1D quasi-parallel shock}
This section presents the results of 1D quasi-parallel shock simulations. The upstream and downstream shock conditions are first determined using the Rankine-Hugoniot relations from the ideal MHD equations, with an assumed electron-to-ion pressure ratio $p_e/p_i = 2$ to specify the electron and ion pressures. While this assumption is reasonable for the upstream region, it does not accurately reflect the downstream, where the ion pressure typically exceeds the electron pressure by a factor of 5--10. To obtain more realistic downstream conditions, we perform an MHD simulation that solves the ideal MHD equations along with an independent electron pressure equation. This approach provides improved initial conditions for the PIC simulations, although the Rankine-Hugoniot jump conditions alone already offer a good starting point.

\begin{figure}
  \centering
  \includegraphics[width=1.0\textwidth]{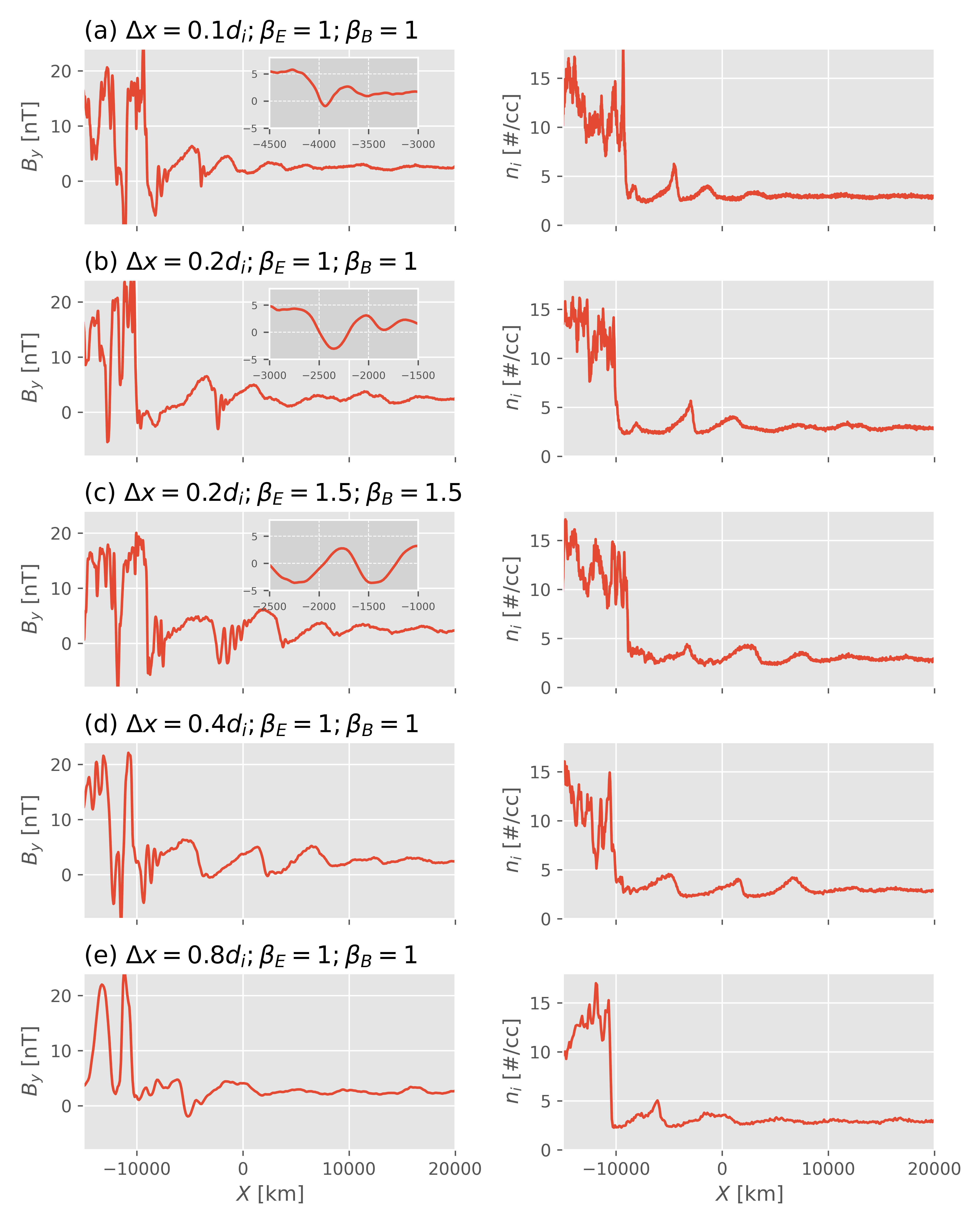}
  \caption{1D quasi-parallel shock simulation results at $t=260~\mathrm{s}$ for different grid resolutions and flux limiter parameters $\beta$. }
  \label{fig:shock30deg}
\end{figure}

The angle between the shock normal and the magnetic field is set to $30^\circ$.
With $p_e/p_i=2$, the upstream conditions used in this test are:
\begin{equation}
\begin{bmatrix}
\rho \\
u_x \\
u_y \\
u_z \\
B_x \\
B_y \\
B_z \\
p_e \\
p_i
\end{bmatrix}
=
\begin{bmatrix}
3~\mathrm{amu/cc} \\
-440.6~\mathrm{km/s} \\
0 \\
0 \\
4.33~\mathrm{nT} \\
2.5~\mathrm{nT} \\
0 \\
0.00995~\mathrm{nPa} \\
0.004975~\mathrm{nPa}
\end{bmatrix},
\end{equation}
and the downstream conditions are:
\begin{equation}
\begin{bmatrix}
\rho \\
u_x \\
u_y \\
u_z \\
B_x \\
B_y \\
B_z \\
p_e \\
p_i
\end{bmatrix}
=
\begin{bmatrix}
10.9~\mathrm{amu/cc} \\
-121.3~\mathrm{km/s} \\
-10.86~\mathrm{km/s} \\
0 \\
4.33~\mathrm{nT} \\
9.47~\mathrm{nT} \\
0 \\
0.4588~\mathrm{nPa} \\
0.2294~\mathrm{nPa}
\end{bmatrix}.
\end{equation}
The corresponding upstream ion inertial length is $d_i=133~\mathrm{km}$, the Alfv\'en Mach number is $M_\text{A}=7$, and the ion gyroperiod is $2\pi/\omega_{ci}=13~\mathrm{s}$. The ion-electron mass ratio is set to $m_i/m_e=25$, and a reduced light speed of $c=3000~\mathrm{km/s}$ is used. The simulation domain spans $[-1.8\times10^4, 3.2\times10^4]~\mathrm{km}$, with the initial shock front located at $x=0$. The CFL number is 0.2. Simulation results at $t=260~\mathrm{s}$ for different grid resolutions and flux limiter parameters $\beta$ are shown in Figure~\ref{fig:shock30deg}. All simulations employ the Lax-Friedrichs diffusion with a flux limiter, compute the current in the comoving frame, and apply one pass of digital filter smoothing to the current density. Without these techniques, the simulation becomes numerically unstable for this test.

Panels a, b, d, and e display results for varying grid resolutions. Simulations with $\Delta x \le 0.4d_i$ accurately resolve the upstream fast magnetosonic waves, while $\Delta x = 0.8d_i$ introduces excessive numerical diffusion, resulting in poor wave resolution. The fast magnetosonic waves steepen and form shocklets, and short-wavelength oscillations are generated at the leading edge of these shocklets. Similar wave structures have been observed in Earth's foreshock region. The wavelengths in the simulations with $\Delta x = 0.1d_i$ and $\Delta x = 0.2d_i$ are nearly identical, indicating that these are likely physical, rather than numerical, waves. Panel (c) shows results with a flux limiter parameter $\beta = 1.5$, which is less diffusive and produces short-wavelength oscillations with larger amplitude, but the overall solution remains similar to that with $\beta = 1$.

The grid convergence study demonstrates that the method can resolve upstream fast magnetosonic waves with a grid size of $\Delta x = 0.4d_i$ or smaller. The close agreement among these simulations indicates that the solution is converged.

To further validate the numerical methods, Zhou et al. (2025) \cite{zhou2025planar} conduct a comprehensive suite of 1D and 2D planar shock simulations using a similar setup. It provides a detailed analysis of the resulting particle distributions and electromagnetic field structures generated by the shock, offering additional verification of the accuracy and physical fidelity of the proposed algorithms.

\subsection{1D quasi-perpendicular shock}
For a quasi-perpendicular shock, reflected particles are unable to penetrate far into the upstream region, and fast magnetosonic waves are not generated. However, whistler-mode waves can be excited at the shock front and propagate upstream under suitable conditions. This 1D quasi-perpendicular shock problem is designed to test the generation and propagation of whistler waves.

\begin{figure}
 \centering
\includegraphics[width=0.9\textwidth]{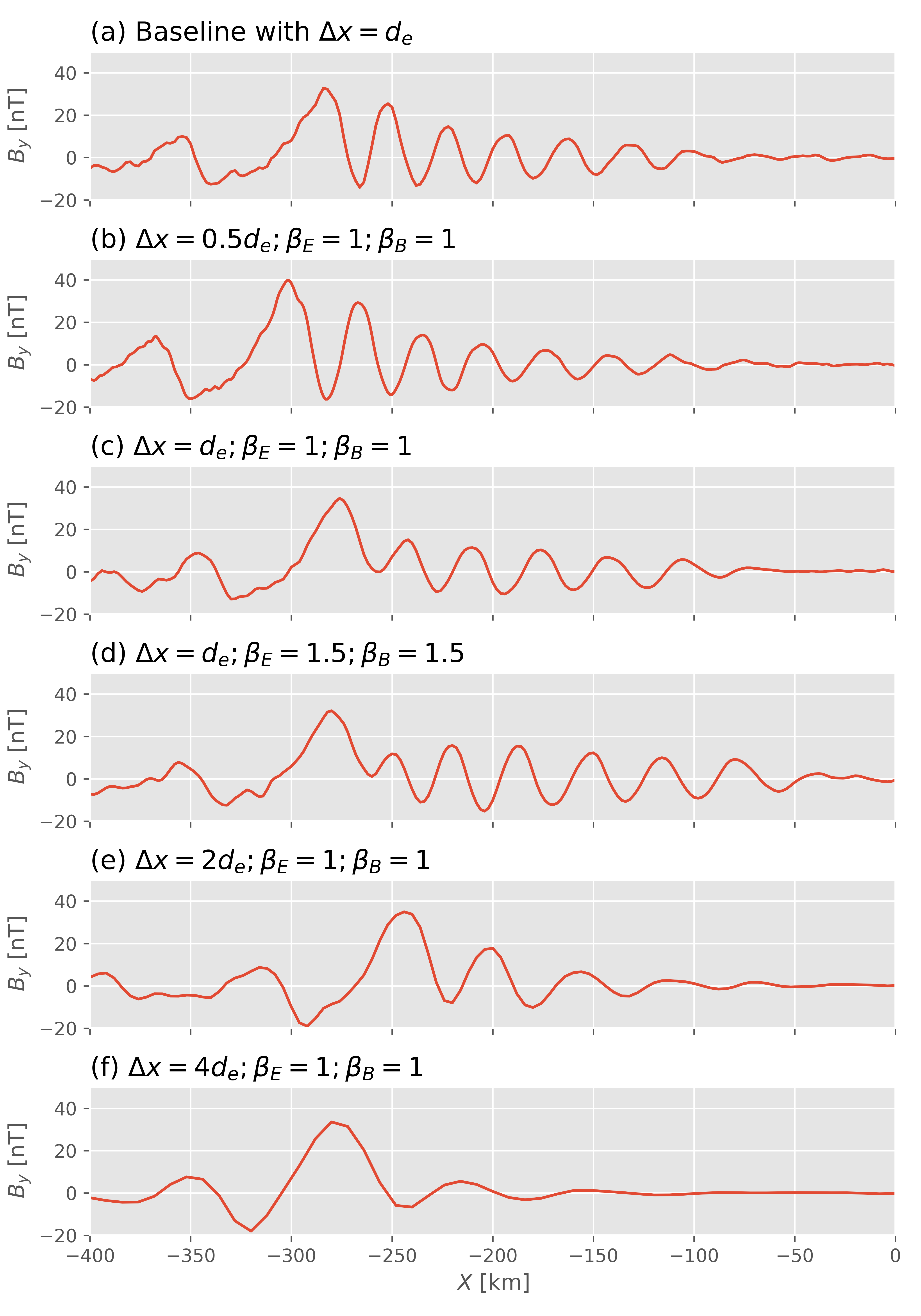}
  \caption{1D quasi-perpendicular shock simulation results at $t=6~\mathrm{s}$ for different grid resolutions parameters. The baseline case (panel a) does not include the Lax-Friedrichs diffusion and current noise reduction methods. Panels b-f include these techniques, and the current density is smoothed once.}
   \label{fig:shock60deg}
\end{figure}

Because whistler waves are frequently observed in Mercury's foreshock region, the initial conditions are chosen to be representative of the solar wind at Mercury. The angle between the shock normal and the magnetic field is set to $60^\circ$ in this test. The initial conditions are determined using the same procedure as in the quasi-parallel shock test, but with an initial $p_e/p_i$ ratio of 1. The upstream conditions are:
\begin{equation}
\begin{bmatrix}
\rho \\
u_x \\
u_y \\
u_z \\
B_x \\
B_y \\
B_z \\
p_e \\
p_i
\end{bmatrix}
=
\begin{bmatrix}
40~\mathrm{amu/cc} \\
-552~\mathrm{km/s} \\
0 \\
0 \\
20~\mathrm{nT} \\
0 \\
34.6~\mathrm{nT} \\
0.27613~\mathrm{nPa} \\
0.27613~\mathrm{nPa}
\end{bmatrix},
\end{equation}
and the downstream conditions are:
\begin{equation}
\begin{bmatrix}
\rho \\
u_x \\
u_y \\
u_z \\
B_x \\
B_y \\
B_z \\
p_e \\
p_i
\end{bmatrix}
=
\begin{bmatrix}
119.5~\mathrm{amu/cc} \\
-184.9~\mathrm{km/s} \\
0 \\
-31.15~\mathrm{km/s} \\
20~\mathrm{nT} \\
0 \\
106.85~\mathrm{nT} \\
5.02~\mathrm{nPa} \\
5.02~\mathrm{nPa}
\end{bmatrix}.
\end{equation}
The ion-electron mass ratio is set to $m_i/m_e=400$, and a reduced light speed of $c=10,000~\mathrm{km/s}$ is used. The upstream ion and electron inertial lengths are $d_i=36~\mathrm{km}$ and $d_e = 1.8~\mathrm{km}$, respectively. The Alfv\'en Mach number is $M_\text{A}=4$, and the ion gyroperiod is $2\pi/\omega_{ci}=1.64~\mathrm{s}$. The simulation domain spans $[-2\times10^3, 2\times10^3]~\mathrm{km}$, with the initial shock front located at $x=0$. The CFL number is 0.2. Simulation results at $t=6~\mathrm{s}$ for different grid resolutions and flux limiter parameters $\beta$ are shown in Figure~\ref{fig:shock60deg}.

Even without applying the Lax-Friedrichs diffusion and the current noise reduction method, the simulation remains numerically stable, as demonstrated by the baseline case in panel a. All other simulations employ the Lax-Friedrichs diffusion with a flux limiter, compute the current in the comoving frame, and apply one pass of current smoothing. The results with (panel a) and without (panels c and d) these techniques exhibit similar wave amplitudes and wavelengths, indicating that the numerical methods introduced in this paper do not adversely affect the physical solution. The simulation with $\beta=1.5$ (panel d) preserves the whistler wave amplitudes better than the case with $\beta=1$ (panel c), as expected due to the reduced numerical diffusion at higher $\beta$. The grid convergence study (panels b, c, e, and f) demonstrates that, even with a coarse grid size of $\Delta x = 4d_e$, whistler waves are still generated and propagate upstream, although their amplitudes are rapidly damped. When the cell size is reduced to $d_e$ or smaller, the simulations produce consistent wave amplitudes and wavelengths.
 
\subsection{Magnetic reconnection}
\begin{table}[H]
  \centering
  \caption{Parameters of the magnetic reconnection simulations.}
  \begin{tabular}{|c|c|c|c|c|c|}
    \hline
    ID & Comoving current & LF diffusion & Flux limiter $\beta$ & ppc & $ns_j$   \\
    \hline
    (a) & No  & No & N/A & 1600 & 0 \\
    (b) & No  & No & N/A &  100 & 0 \\
    (c) & Yes & Yes & 1   &  100 & 0 \\
    (d) & Yes & Yes & 1   &  100 & 1 \\
    (e) & Yes & Yes & 1.5 &  100 & 1 \\
    (f) & Yes & Yes & 1   &  100 & 3 \\
    \hline
    \multicolumn{6}{l}{* $ns_j$ is the number of passes of smoothing applied to the current.} \\
  \end{tabular}
  \label{tab:mr}
\end{table}

\begin{figure}
  \centering
  \includegraphics[width=1.0\textwidth]{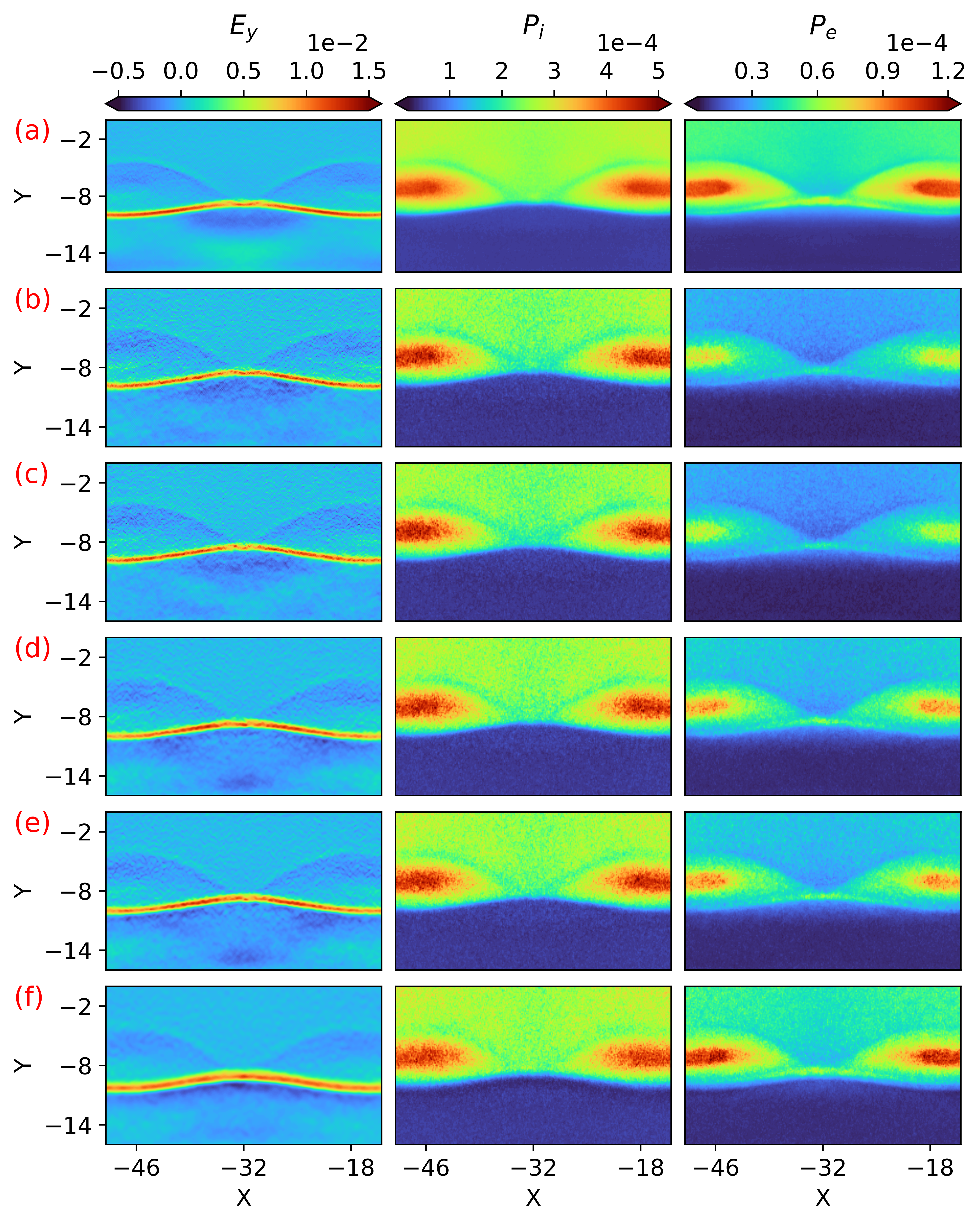}
  \caption{Magnetic reconnection simulation results for simulations listed in Table~\ref{tab:mr}.}
  \label{fig:mr}
\end{figure}

The original GL-ECSIM method is robust and accurate for magnetic reconnection simulations. To assess the impact of the new numerical techniques introduced in this work, we perform a series of 2D asymmetric magnetic reconnection simulations using the same setup as in \cite{chen_fleks_2023}. The simulation parameters for each case are summarized in Table~\ref{tab:mr}. Simulations (a) and (b) serve as baseline cases without the new techniques. Simulation (a) uses a high number of particles per cell (ppc=1600) to minimize statistical noise, while simulation (b) uses a lower particle count (ppc=100) as other simulations listed in the table.

Simulations (c) through (f) incorporate the Lax-Friedrichs (LF) diffusion with a flux limiter and the comoving frame current calculation. Comparing simulations (b) and (c) demonstrates that the inclusion of the LF diffusion and comoving frame current has a negligible effect on the results. 

The effect of current smoothing is examined in simulations (d)–(f). Applying one or more passes of digital filter smoothing to the current density help to preserve the electron pressure $p_e$ in the reconnection exhaust, indicating a reduction in numerical cooling. Increasing the flux limiter parameter from $\beta=1$ (simulation (d)) to $\beta=1.5$ (simulation (e)) produces only minor differences.

In summary, these simulations demonstrate that:
\begin{itemize}
  \item The application of limited Lax-Friedrichs diffusion and comoving frame current calculation does not adversely affect the accuracy or physical fidelity of magnetic reconnection simulations.
  \item The reconnection results are insensitive to the choice of flux limiter parameter $\beta$.
  \item Smoothing the current density reduces numerical cooling of electrons.
\end{itemize}


\section{Summary}
In this paper, we enhance the GL-ECSIM particle-in-cell (PIC) method for shock simulations by introducing two key numerical techniques: (1) incorporating a Lax-Friedrichs-type diffusion term with a flux limiter into the Maxwell solver to effectively suppress spurious oscillations at discontinuities, and (2) computing the current density in the comoving frame to reduce statistical particle noise. While these algorithms are implemented and validated within a semi-implicit PIC code, they are applicable to explicit PIC schemes as well.

Comprehensive numerical tests, including electromagnetic wave propagation with discontinuities, uniform fast flow, collisionless shock, and asymmetric magnetic reconnection simulations, demonstrate the accuracy and stability of the proposed methods. The Lax-Friedrichs diffusion with a flux limiter efficiently eliminates nonphysical oscillations, and the comoving frame current calculation substantially reduces noise and stabilizes simulations of fast plasma flows. Additionally, applying a digital filter to the current density further enhances numerical stability and mitigates numerical cooling effects.

 These improvements enhance the robustness and fidelity of the semi-implicit PIC code FLEKS, enabling reliable simulations of complex, three-dimensional planetary shock structures. 
 These simulations will be used to investigate the role of the kinetic shocks in magnetospheres \cite{Zou2025}.

\section*{Acknowledgment}

The authors acknowledge support from the NASA grants 80NSSC24K0144 and 80NSSC23K1409. Computational resources were provided by the NASA High-End Computing (HEC) Program through the NASA Advanced Supercomputing (NAS) Division at Ames Research Center.

\bibliographystyle{elsarticle-num-names}
\bibliography{shock,csem}

\begin{thebibliography}{24}
\expandafter\ifx\csname natexlab\endcsname\relax\def\natexlab#1{#1}\fi
\providecommand{\url}[1]{\texttt{#1}}
\providecommand{\href}[2]{#2}
\providecommand{\path}[1]{#1}
\providecommand{\DOIprefix}{doi:}
\providecommand{\ArXivprefix}{arXiv:}
\providecommand{\URLprefix}{URL: }
\providecommand{\Pubmedprefix}{pmid:}
\providecommand{\doi}[1]{\href{http://dx.doi.org/#1}{\path{#1}}}
\providecommand{\Pubmed}[1]{\href{pmid:#1}{\path{#1}}}
\providecommand{\bibinfo}[2]{#2}
\ifx\xfnm\relax \def\xfnm[#1]{\unskip,\space#1}\fi
\bibitem[{Taflove et~al.(2005)Taflove, Hagness, and Piket-May}]{taflove2005computational}
\bibinfo{author}{A.~Taflove}, \bibinfo{author}{S.~C. Hagness}, \bibinfo{author}{M.~Piket-May},
\newblock \bibinfo{title}{Computational electromagnetics: the finite-difference time-domain method},
\newblock \bibinfo{journal}{The Electrical Engineering Handbook} \bibinfo{volume}{3} (\bibinfo{year}{2005}) \bibinfo{pages}{15}.
\bibitem[{Birdsall and Langdon(2018)}]{birdsall2018plasma}
\bibinfo{author}{C.~K. Birdsall}, \bibinfo{author}{A.~B. Langdon}, \bibinfo{title}{Plasma physics via computer simulation}, \bibinfo{publisher}{CRC press}, \bibinfo{year}{2018}.
\bibitem[{Fi{\'u}za et~al.(2011)Fi{\'u}za, Marti, Fonseca, Silva, Tonge, May, and Mori}]{fiuza2011efficient}
\bibinfo{author}{F.~Fi{\'u}za}, \bibinfo{author}{M.~Marti}, \bibinfo{author}{R.~Fonseca}, \bibinfo{author}{L.~Silva}, \bibinfo{author}{J.~Tonge}, \bibinfo{author}{J.~May}, \bibinfo{author}{W.~B. Mori},
\newblock \bibinfo{title}{Efficient modeling of laser--plasma interactions in high energy density scenarios},
\newblock \bibinfo{journal}{Plasma Physics and Controlled Fusion} \bibinfo{volume}{53} (\bibinfo{year}{2011}) \bibinfo{pages}{074004}.
\bibitem[{Arber et~al.(2015)Arber, Bennett, Brady, Lawrence-Douglas, Ramsay, Sircombe, Gillies, Evans, Schmitz, Bell, and Ridgers}]{arber_contemporary_2015}
\bibinfo{author}{T.~D. Arber}, \bibinfo{author}{K.~Bennett}, \bibinfo{author}{C.~S. Brady}, \bibinfo{author}{A.~Lawrence-Douglas}, \bibinfo{author}{M.~G. Ramsay}, \bibinfo{author}{N.~J. Sircombe}, \bibinfo{author}{P.~Gillies}, \bibinfo{author}{R.~G. Evans}, \bibinfo{author}{H.~Schmitz}, \bibinfo{author}{A.~R. Bell}, \bibinfo{author}{C.~P. Ridgers},
\newblock \bibinfo{title}{Contemporary particle-in-cell approach to laser-plasma modelling},
\newblock \bibinfo{journal}{Plasma Physics and Controlled Fusion} \bibinfo{volume}{57} (\bibinfo{year}{2015}) \bibinfo{pages}{113001}. \DOIprefix\doi{10.1088/0741-3335/57/11/113001}.
\bibitem[{Lapenta(2017)}]{lapenta_exactly_2017}
\bibinfo{author}{G.~Lapenta},
\newblock \bibinfo{title}{Exactly energy conserving semi-implicit particle in cell formulation},
\newblock \bibinfo{journal}{Journal of Computational Physics} \bibinfo{volume}{334} (\bibinfo{year}{2017}) \bibinfo{pages}{349--366}. \URLprefix \url{http://dx.doi.org/10.1016/j.jcp.2017.01.002}. \DOIprefix\doi{10.1016/j.jcp.2017.01.002}, \bibinfo{note}{arXiv: 1602.06326 Publisher: Elsevier Inc.}
\bibitem[{Vay et~al.(2011)Vay, Geddes, Cormier-Michel, and Grote}]{vay_numerical_2011}
\bibinfo{author}{J.-L. Vay}, \bibinfo{author}{C.~Geddes}, \bibinfo{author}{E.~Cormier-Michel}, \bibinfo{author}{D.~Grote},
\newblock \bibinfo{title}{Numerical methods for instability mitigation in the modeling of laser wakefield accelerators in a {Lorentz}-boosted frame},
\newblock \bibinfo{journal}{Journal of Computational Physics} \bibinfo{volume}{230} (\bibinfo{year}{2011}) \bibinfo{pages}{5908--5929}. \URLprefix \url{https://linkinghub.elsevier.com/retrieve/pii/S0021999111002270}. \DOIprefix\doi{10.1016/j.jcp.2011.04.003}.
\bibitem[{Werner et~al.(2025)Werner, Adams, and Cary}]{werner2025suppressing}
\bibinfo{author}{G.~R. Werner}, \bibinfo{author}{L.~C. Adams}, \bibinfo{author}{J.~R. Cary},
\newblock \bibinfo{title}{Suppressing grid instability and noise in particle-in-cell simulation by smoothing},
\newblock \bibinfo{journal}{arXiv preprint arXiv:2503.05123}  (\bibinfo{year}{2025}).
\bibitem[{Chen et~al.(2023)Chen, Tóth, Zhou, and Wang}]{chen_fleks_2023}
\bibinfo{author}{Y.~Chen}, \bibinfo{author}{G.~Tóth}, \bibinfo{author}{H.~Zhou}, \bibinfo{author}{X.~Wang},
\newblock \bibinfo{title}{{FLEKS}: {A} flexible particle-in-cell code for multi-scale plasma simulations},
\newblock \bibinfo{journal}{Computer Physics Communications} \bibinfo{volume}{287} (\bibinfo{year}{2023}) \bibinfo{pages}{108714}. \URLprefix \url{https://www.sciencedirect.com/science/article/pii/S0010465523000590}. \DOIprefix\doi{10.1016/j.cpc.2023.108714}.
\bibitem[{Chen and Tóth(2019)}]{Chen2019a}
\bibinfo{author}{Y.~Chen}, \bibinfo{author}{G.~Tóth},
\newblock \bibinfo{title}{Gauss's {Law} satisfying {Energy}-{Conserving} {Semi}-{Implicit} {Particle}-in-{Cell} method},
\newblock \bibinfo{journal}{Journal of Computational Physics} \bibinfo{volume}{386} (\bibinfo{year}{2019}) \bibinfo{pages}{632--652}. \URLprefix \url{https://doi.org/10.1016/j.jcp.2019.02.032}. \DOIprefix\doi{10.1016/j.jcp.2019.02.032}, \bibinfo{note}{arXiv: 1808.05745 Publisher: Elsevier Inc.}
\bibitem[{Hirsch(1989)}]{Hirsch:1989}
\bibinfo{author}{C.~Hirsch}, \bibinfo{title}{Numerical Computation of Internal and External Flows, Volume 1, Fundamentals of Numerical Discretization}, \bibinfo{publisher}{John Wiley \& Sons}, \bibinfo{address}{Toronto}, \bibinfo{year}{1989}.
\bibitem[{Harten and Zwas(1972)}]{Harten:1972}
\bibinfo{author}{A.~Harten}, \bibinfo{author}{G.~Zwas},
\newblock \bibinfo{title}{Self-adjusting hybrid schemes for shock computation},
\newblock \bibinfo{journal}{J. Comput. Phys.} \bibinfo{volume}{9} (\bibinfo{year}{1972}) \bibinfo{pages}{568--583}.
\bibitem[{{van Leer}(1979)}]{vanLeer:1979}
\bibinfo{author}{B.~{van Leer}},
\newblock \bibinfo{title}{Towards the ultimate conservative difference scheme. {V}. {A} second-order sequel to {G}odunov's method},
\newblock \bibinfo{journal}{J. Comput. Phys.} \bibinfo{volume}{32} (\bibinfo{year}{1979}) \bibinfo{pages}{101--136}.
\bibitem[{LeVeque(2002)}]{leveque2002finite}
\bibinfo{author}{R.~J. LeVeque}, \bibinfo{title}{Finite volume methods for hyperbolic problems}, volume~\bibinfo{volume}{31}, \bibinfo{publisher}{Cambridge university press}, \bibinfo{year}{2002}.
\bibitem[{Yee(1966)}]{yee1966numerical}
\bibinfo{author}{K.~Yee},
\newblock \bibinfo{title}{Numerical solution of initial boundary value problems involving maxwell's equations in isotropic media},
\newblock \bibinfo{journal}{IEEE Transactions on antennas and propagation} \bibinfo{volume}{14} (\bibinfo{year}{1966}) \bibinfo{pages}{302--307}.
\bibitem[{Lapenta(2017)}]{Lapenta:2017}
\bibinfo{author}{G.~Lapenta},
\newblock \bibinfo{title}{Exactly energy conserving semi-implicit particle in cell formulation},
\newblock \bibinfo{journal}{J. Comput. Phys.} \bibinfo{volume}{334} (\bibinfo{year}{2017}) \bibinfo{pages}{349}. \DOIprefix\doi{10.1016/j.jcp.2017.01.002}.
\bibitem[{Dedner et~al.(2003)Dedner, Kemm, Kr{\"o}ner, Munz, Schnitzer, and Wesenberg}]{Dedner:2001}
\bibinfo{author}{A.~Dedner}, \bibinfo{author}{F.~Kemm}, \bibinfo{author}{D.~Kr{\"o}ner}, \bibinfo{author}{C.~Munz}, \bibinfo{author}{T.~Schnitzer}, \bibinfo{author}{M.~Wesenberg},
\newblock \bibinfo{title}{Hyperbolic divergence cleaning for the {MHD} equations},
\newblock \bibinfo{journal}{J. Comput. Phys.} \bibinfo{volume}{175} (\bibinfo{year}{2003}) \bibinfo{pages}{645--673}.
\bibitem[{Tóth et~al.(2012)Tóth, van~der Holst, Sokolov, De~Zeeuw, Gombosi, Fang, Manchester, Meng, Najib, Powell, Stout, Glocer, Ma, and Opher}]{toth_adaptive_2012}
\bibinfo{author}{G.~Tóth}, \bibinfo{author}{B.~van~der Holst}, \bibinfo{author}{I.~V. Sokolov}, \bibinfo{author}{D.~L. De~Zeeuw}, \bibinfo{author}{T.~I. Gombosi}, \bibinfo{author}{F.~Fang}, \bibinfo{author}{W.~B. Manchester}, \bibinfo{author}{X.~Meng}, \bibinfo{author}{D.~Najib}, \bibinfo{author}{K.~G. Powell}, \bibinfo{author}{Q.~F. Stout}, \bibinfo{author}{A.~Glocer}, \bibinfo{author}{Y.~J. Ma}, \bibinfo{author}{M.~Opher},
\newblock \bibinfo{title}{Adaptive numerical algorithms in space weather modeling},
\newblock \bibinfo{journal}{Journal of Computational Physics} \bibinfo{volume}{231} (\bibinfo{year}{2012}) \bibinfo{pages}{870--903}. \DOIprefix\doi{10.1016/j.jcp.2011.02.006}.
\bibitem[{Chen et~al.(2017)Chen, Tóth, Cassak, Jia, Gombosi, Slavin, Markidis, Peng, Jordanova, and Henderson}]{chen_global_2017}
\bibinfo{author}{Y.~Chen}, \bibinfo{author}{G.~Tóth}, \bibinfo{author}{P.~Cassak}, \bibinfo{author}{X.~Jia}, \bibinfo{author}{T.~I. Gombosi}, \bibinfo{author}{J.~A. Slavin}, \bibinfo{author}{S.~Markidis}, \bibinfo{author}{I.~B. Peng}, \bibinfo{author}{V.~K. Jordanova}, \bibinfo{author}{M.~G. Henderson},
\newblock \bibinfo{title}{Global {Three}-{Dimensional} {Simulation} of {Earth}'s {Dayside} {Reconnection} {Using} a {Two}-{Way} {Coupled} {Magnetohydrodynamics} {With} {Embedded} {Particle}-in-{Cell} {Model}: {Initial} {Results}},
\newblock \bibinfo{journal}{Journal of Geophysical Research: Space Physics} \bibinfo{volume}{122} (\bibinfo{year}{2017}) \bibinfo{pages}{10,318--10,335}. \DOIprefix\doi{10.1002/2017JA024186}.
\bibitem[{Chen et~al.(2020)Chen, Tóth, Hietala, Vines, Zou, Nishimura, Silveira, Guo, Lin, and Markidis}]{Chen2020}
\bibinfo{author}{Y.~Chen}, \bibinfo{author}{G.~Tóth}, \bibinfo{author}{H.~Hietala}, \bibinfo{author}{S.~K. Vines}, \bibinfo{author}{Y.~Zou}, \bibinfo{author}{Y.~Nishimura}, \bibinfo{author}{M.~V. Silveira}, \bibinfo{author}{Z.~Guo}, \bibinfo{author}{Y.~Lin}, \bibinfo{author}{S.~Markidis},
\newblock \bibinfo{title}{Magnetohydrodynamic with embedded particle-in-cell simulation of the {Geospace} {Environment} {Modeling} dayside kinetic processes challenge event},
\newblock \bibinfo{journal}{arXiv}  (\bibinfo{year}{2020}). \DOIprefix\doi{10.1029/2020ea001331}.
\bibitem[{Chen et~al.(2019)Chen, Tóth, Jia, Slavin, Sun, Markidis, Gombosi, and Raines}]{Chen2019}
\bibinfo{author}{Y.~Chen}, \bibinfo{author}{G.~Tóth}, \bibinfo{author}{X.~Jia}, \bibinfo{author}{J.~A. Slavin}, \bibinfo{author}{W.~Sun}, \bibinfo{author}{S.~Markidis}, \bibinfo{author}{T.~I. Gombosi}, \bibinfo{author}{J.~M. Raines},
\newblock \bibinfo{title}{Studying {Dawn}-{Dusk} {Asymmetries} of {Mercury}'s {Magnetotail} {Using} {MHD}-{EPIC} {Simulations}},
\newblock \bibinfo{journal}{Journal of Geophysical Research: Space Physics} \bibinfo{volume}{124} (\bibinfo{year}{2019}) \bibinfo{pages}{8954--8973}. \DOIprefix\doi{10.1029/2019JA026840}, \bibinfo{note}{arXiv: 1904.06753}.
\bibitem[{Vay et~al.(2002)Vay, Colella, McCORQUODALE, Van~Straalen, Friedman, and Grote}]{vay_mesh_2002}
\bibinfo{author}{J.-L. Vay}, \bibinfo{author}{P.~Colella}, \bibinfo{author}{P.~McCORQUODALE}, \bibinfo{author}{B.~Van~Straalen}, \bibinfo{author}{A.~Friedman}, \bibinfo{author}{D.~Grote},
\newblock \bibinfo{title}{Mesh refinement for particle-in-cell plasma simulations: {Applications} to and benefits for heavy ion fusion},
\newblock \bibinfo{journal}{Laser and Particle Beams} \bibinfo{volume}{20} (\bibinfo{year}{2002}) \bibinfo{pages}{569--575}. \DOIprefix\doi{10.1017/S0263034602204139}.
\bibitem[{Lapenta(2012)}]{lapenta_particle_2012}
\bibinfo{author}{G.~Lapenta},
\newblock \bibinfo{title}{Particle simulations of space weather},
\newblock \bibinfo{journal}{Journal of Computational Physics} \bibinfo{volume}{231} (\bibinfo{year}{2012}) \bibinfo{pages}{795--821}. \URLprefix \url{http://dx.doi.org/10.1016/j.jcp.2011.03.035}. \DOIprefix\doi{10.1016/j.jcp.2011.03.035}, \bibinfo{note}{publisher: Elsevier Inc.}
\bibitem[{Zhou et~al.(2025)Zhou, Chen, Dong, Wang, Zou, Walsh, and Tóth}]{zhou2025planar}
\bibinfo{author}{H.~Zhou}, \bibinfo{author}{Y.~Chen}, \bibinfo{author}{C.~Dong}, \bibinfo{author}{L.~Wang}, \bibinfo{author}{Y.~Zou}, \bibinfo{author}{B.~Walsh}, \bibinfo{author}{G.~Tóth}, \bibinfo{title}{Planar collisionless shock simulations with semi-implicit particle-in-cell model fleks}, \bibinfo{year}{2025}. \href{http://arxiv.org/abs/2506.08384}{{\tt arXiv:2506.08384}}.
\bibitem[{Zou et~al.(2025)Zou, Walsh, Chen, Zhou, and Raptis}]{Zou2025}
\bibinfo{author}{Y.~Zou}, \bibinfo{author}{B.~M. Walsh}, \bibinfo{author}{Y.~Chen}, \bibinfo{author}{H.~Zhou}, \bibinfo{author}{S.~Raptis},
\newblock \bibinfo{title}{Control of solar wind on magnetic field fluctuations in the subsolar magnetosheath},
\newblock \bibinfo{journal}{Journal of Geophysical Research: Space Physics} \bibinfo{volume}{130} (\bibinfo{year}{2025}) \bibinfo{pages}{e2025JA033856}. \DOIprefix\doi{https://doi.org/10.1029/2025JA033856}.

\end{thebibliography}
\end{document}